\documentclass[12pt]{article}
\title{ The QCD spectrum: mixing, strong decays and the role of sea quarks}
\author{Yu.A.Simonov\\
 State Research
Center\\Institute of Theoretical and Experimental Physics, \\
Moscow, 117218 Russia}
 \date{}
\newcommand{\beq}{\begin{eqnarray}}
 \newcommand{\eeq}{\end{eqnarray}}
\newcommand{\be}{\begin{equation}}
 \newcommand{\ee}{\end{equation}}

\def\ga{\mathrel{\mathpalette\fun >}}
\def\fun#1#2{\lower3.6pt\vbox{\baselineskip0pt\lineskip.9pt
\ialign{$\mathsurround=0pt#1\hfil ##\hfil$\crcr#2\crcr\sim\crcr}}}

\newcommand{{\SD}}{\rm SD}

\newcommand{\vex}{\mbox{\boldmath${\rm x}$}}

\newcommand{\ver}{\mbox{\boldmath${\rm r}$}}

\newcommand{\veP}{\mbox{\boldmath${\rm P}$}}
\newcommand{\vep}{\mbox{\boldmath${\rm p}$}}

\newcommand{\vek}{\mbox{\boldmath${\rm k}$}}

\newcommand{\lan}{\langle}
\newcommand{\ran}{\rangle}
\begin{document}
\maketitle

\begin{abstract}
The light hadron spectrum  as computed in nonperturbative QCD is
reviewed and compared to lattice data and experiment. The mixing
of mesons, hybrids and glueballs is calculated in the Field
Correlator Method. The strong decay mechanisms are  found  out in
the  method and compared to the known phenomenological models. The
role of sea quarks (unquenched approximation) is studied
analytically using radially excited mesons as an example, and
compared to experiment.
\end{abstract}

\section{Introduction}

In the last decade there has been a substantial progress in
understanding the QCD spectrum, both in analytical methods
\cite{1}, and lattice calculations \cite{2}. On the analytic side
the most economic and promising turns out  the Field Correlator
Method (FCM) \cite{3}, which starts from the basic principles and
correct description of the  QCD vacuum with the help of
gauge-invariant field correlators. It was proved \cite{4} that
confinement can be described by the  lowest, quadratic correlator
$\lan F_{\mu\nu} (x) F_{\lambda\sigma} (y) \ran $ which contains
two scalar formfactors $D(x-y) $ and $D_1 (x-y)$. Lattice data
\cite{5} show that $D$ and $D_1$   describe $Q\bar Q$ static
potential with a few percent accuracy, and the string tension
$\sigma$ is obtained directly from $D(x)$: $\sigma=\frac12 \int
D(x) d^2x$.

In addition $D(x) \equiv D(x/T_g)$ contains another important
parameter  -- the gluon correlation length $T_g$, which was found
on the lattice \cite{6} and analytically \cite{7} to be very
small, $T_g \leq 0.2$ fm.

This circumstance allows to develop the local Hamiltonian and
Lagrangian methods for the description of the $q\bar q$  and $3q$
bound states, which will be discussed in the next section.

However, this is not the whole story since there are valence
gluons which can also form bound states by themselves (glueballs)
and with the $q\bar q$ (hybrids). To introduce these states one
should define the valence gluons in contrast to the
nonperturbative background, and one can do it unambigiously in the
framework of the background perturbation theory  \cite{8}. In this
way one gets the  local Hamiltonian  also for hybrids \cite{9,10}
and for glueballs \cite{11} and calculate the corresponding
spectra in good agreement with lattice data. In doing so one
realizes that a gluon excitation "costs" around 1 GeV, which
allows  to disregard these states in the first approximation  when
computing the lowest meson or baryon states.

However the exact treatment requires the introduction of Fock
tower of states, and consideration of mixing between meson and
hybrid states, which is done in Section 3.  The effects of sea
quarks on the spectrum   and strong decays is considered in
section 4, while Section 5 is devoted to conclusions.

\section{Hamiltonian}

There are two possible approaches to incorporating nonperturbative
field correlators  in the quark-antiquark (or $3q$) dynamics. The
first has to deal with the effective nonlocal quark Lagrangian
containing field correlators \cite{12}. From this one obtains
first-order Dirac-type  integro-differential equations for
heavy-light  mesons \cite{12,13}, light mesons  and baryons
\cite{14}. These equations contain the effect of chiral symmetry
breaking \cite{12} which is directly connected to confinement.

The second approach is based on the effective Hamiltonian for any
gauge-invariant quark-gluon system. In the limit $T_g\to 0$ this
Hamiltoniam is simple and local, and in most cases when spin
interaction can be considered as a perturbation one obtains
results for the spectra in an analytic form, which is transparent.

For this reason we choose below the second, Hamiltonian approach
\cite{15,16}. We start with the exact Fock-Feynman-Schwinger
Representation for the $q\bar q$ Green's function (for a review
see \cite{17}), taking for simplicity nonzero flavor case $$
 G^{(x,y)}_{q\bar q} =\int^\infty_0 ds_1 \int^\infty_0 ds_2
(Dz)_{xy}(D\bar z)_{xy}e^{-K_1-K_2}\times $$\be\lan  tr
\Gamma_{in} (m_1-\hat D_1) W_\sigma (C) \Gamma_{out} (m_2-\hat
D_2)\ran_A\label{1}\ee
 where $K_i=\int^{s_1}_0 d\tau_i (m_i+\frac14 (\dot
 z_\mu^{(i)})^2), $ $\Gamma_{in,out}=1, \gamma_5,...$ are meson
 vertices, and $W_\sigma(C)$ is the Wilson loop with spin
 insertions, taken along the contour $C$ formed by paths
 $(Dz)_{xy}$ and $(D\bar z)_{xy}$,
 $$
 W_\sigma (C) =P_F P_A\exp (ig \int_CA_\mu dz_\mu)\times$$
 \be
\times \exp (g\int^{s_1}_0
 \sigma^{(1)}_{\mu\nu} F_{\mu\nu} d\tau_1-g\int^{s_2}_0
 \sigma_{\mu\nu}^{(2)}F_{\mu\nu} d\tau_2).\label{2}\ee
 The last factor in (\ref{2}) defines the spin interaction of
 quark and antiquark. The average $\lan W_\sigma\ran_A$ in
 (\ref{1}) can be computed exactly through field correlators $\lan
 F(1)...F(n)\ran_A$, and keeping only the lowest  one,$\lan F(1)
 F(2)\ran$, which yields according to lattice calculation
 \cite{5} accuracy around 1\% \cite{4},
 one obtains
$$ \lan  W_\sigma (C)\ran_A \simeq \exp (-\frac12
[\int_{S_{min}}ds_{\mu\nu}(1)
\int_{S_{min}}ds_{\lambda\sigma}(2)+$$ \be+ \sum^2_{i,j=1}
\int^{s_i}_0
 \sigma^{(i)}_{\mu\nu} d\tau_i\int^{s_j}_0
 \sigma_{\lambda\sigma}^{(j)} d\tau_j]\lan F_{\mu\nu}(1) F_{\lambda\sigma}(2)\ran ).
 \label{3}\ee

Here $d s_{\mu\nu}(i)$ is the  surface element, $i=1,2$, and  the
double integration in (\ref{3}) is performed over the minimal area
surface $S_{min}$ inside the contour $C$.

 The Gaussian correlator $\lan F_{\mu\nu} (1) F_{\lambda\sigma}(2)
 \ran \equiv D_{\mu\nu,\lambda\sigma} (1,2)$ can be rewritten
 identically in terms of two scalar functions $D(x)$ and $D_1(x)$
 \cite{3}, which have been computed on the lattice \cite{6} to
 have the exponential form $D(x), D_1(x) \sim \exp (-|x|/T_g)$ with
 the gluon correlation length $T_g\approx 0.2 $ fm.

As the next step one introduces the einbein variables $\mu_i$ and
$\nu$; the first one to transform the proper times $s_i, \tau_i$
into the actual (Euclidean) times $t_i\equiv z^{(i)}_4$. One has
\cite{16}
\be
2\mu_i(t_i) =\frac{dt_i}{d\tau_i},~~ \int^\infty_0
ds_i(D^4z^{(i)})_{xy} =const \int
D\mu_i(t_i)(D^3z^{(i)})_{xy}.\label{5}\ee The variable $\nu$
enters in the Gaussian representation of the Nambu-Goto form for
$S_{min}$ and its stationary value $\nu_0$ has the physical
meaning of the energy density along the string. In case of several
strings, as in the baryon case or the hybrid case, each piece of
string has its own parameter $\nu^{(i)}.$

To get rid of the path integration in (\ref{1}) one can go over to
the effective Hamiltonian using the identity
\be
G_{q\bar q} (x,y) =\lan x| \exp (-HT) |y\ran\label{6}\ee where $T$
is the evolution parameter corresponding to the hypersurface
chosen for the Hamiltonian: it is the hyperplane $z_4=const$ in
the c.m. case \cite{16}.

The final form of the c.m. Hamiltonian (apart from the spin and
perturbative terms to be discussed later) for  the $q\bar q$ case
is \cite{16,18} $$ H_0=\sum^2_{i=1} \left(
\frac{m^2_i+\vep^2_i}{2\mu_i} +\frac{\mu_i}{2}\right) + \frac{\hat
L^2/r^2}{2[\mu_1(1-\zeta)^2+\mu_2\zeta^2 + \int^1_0 d \beta (\beta
-\zeta)^2\nu (\beta)]}+$$
\be
+\frac{\sigma^2 r^2}{2} \int^1_0 \frac{d\beta}{\nu(\beta)} +
\int^1_0\frac{\nu(\beta)}{2}d\beta.\label{7}\ee

Here $\zeta= (\mu_1+\int_0\beta \nu d\beta)/ (\mu_1+\mu_2+\int^1_0
\beta \nu d \beta)$ and $\mu_i$ and $\nu(\beta)$ are to be found
from the stationary point of the Hamiltonian
\be
\frac{\partial H_0}{\partial\mu_i}|_{\mu_i=\mu_i^{(0)}} =0,~~
\frac{\partial H_0}{\partial\nu}|_{\nu=\nu^{(0)}} =0.\label{8}\ee

Note that $H_0$ contains as input only $m_1, m_2$ and $\sigma$,
where $m_i$ are current masses defined at the scale 1 GeV. The
further analysis is simplified by the observation that for $L=0$
one finds $\nu^{(0)}=\sigma r$ from (\ref{8}) and
$\mu_i=\sqrt{m^2+\vep^2}$, hence $H_0$ becomes the usual
Relativistic Quark Model (RQM) Hamiltonian
\be
H_0(L=0)=\sum_{i=1}^2\sqrt{m_i^2+\vep^2} +\sigma r. \label{9}\ee

 But $H_0$ is not the whole story,
one should take into account 3 additional terms $H_{self},
H_{spin}$ and $H_{Coul}$, namely, spin terms in (\ref{3}) which
produce two types of contributions: self-energy correction
\cite{19} \be H_{self}=\sum^2_{i=1} \frac{\Delta
m^2_q(i)}{2\mu_i},~~ \Delta m^2_q=-\frac{4\sigma}{\pi} \eta
(m_i),~~ \eta(0) \cong 1 \div 0.9,\label{14}\ee where $\eta(m_i)$
is a known function of current mass $m_i$ \cite{19}, and
spin-dependent interaction between quark and antiquark $H_{spin}$
\cite{1,20} which is entirely described by the field correlators
$D(x), D_1(x) $, including also the one-gluon exchange part
present in $D_1(x)$.

Finally one should take into account gluon exchange contributions
\cite{8}, which can be divided into the Coulomb part $H_{Coul}
=-\frac43\frac{\alpha_s(r)}{r},$ and $H_{rad}$ including
space-like gluon exchanges and perturbative self-energy
corrections (we shall systematically omit these corrections since
they are small for light quarks to be discussed below). In
addition there are gluon contributions which are nondiagonal in
number of gluons $n_g$  and quarks (till now only the sector
$n_g=0$ was considered) and therefore mixing meson states with
hybrids and glueballs \cite{21}; we call these terms $H_{mix}$ and
refer the reader to \cite{21} and the cited there references for
more discussion. Assembling all terms together one has the
following total Hamiltonian in the limit of large $N_c$ and small
$T_g$:
\be
H=H_0+H_{self} +H_{spin} +H_{Coul}+H_{rad} +H_{mix}.\label{15}\ee

We start with $H_0=H_R+H_{string}$ where $H_{string}$ is the term
proportional to  $\hat L^2$ in (6) and $H_R$ -- all the rest terms
in (6). The eigenvalues $M_0$ of $H_R$ can be given with 1\%
accuracy by \cite{22}
\be
M^2_0\approx 8\sigma L+4\pi\sigma (n+\frac34)\label{16}\ee where
$n$ is the radial quantum number, $n=0,1,2,...$ Remarkably
$M_0\approx 4\mu_0$, and  for $L=n=0$ one has $\mu_0(0,0)=0.35$
GeV for $\sigma =0.18 $ GeV$^2$, and $\mu_0$ is fast increasing
with growing $n$ and $L$. This fact  partly explains that spin
interactions become unimportant beyond $L=0,1,2$ since they are
proportional to $d\tau_1 d\tau_2\sim \frac{1}{4\mu_1\mu_2}
dt_1dt_2$ (see (\ref{3}) and \cite{1,20}). Thus constituent mass
(which is actually "constituent energy") $\mu_0$ is  "running".
The   validity of $\mu_0$ as a socially accepted "constituent
mass" is confirmed by its numerical value given above, the spin
splittings of light and heavy mesons  \cite{23} and by baryon
magnetic moments expressed directly through $\mu_0$, and being in
agreement with experimental values \cite{24}.

   We now come to the gluon-containing systems, hybrids and
   glueballs. Referring the reader to the original papers
   \cite{9}-\cite{11} one can recapitulate the main results for
   the spectrum. In both cases the total Hamiltonian has the same
   form as in (\ref{15}), however the contribution of corrections
   differs.

   For glueballs it was argued in \cite{11} that $H_0$ (\ref{7})
   has the same form, but with $m_i=0$ and $\sigma\to
   \sigma_{adj}=\frac{9}{4} \sigma$ while $H_{self} =0$ due to
   gauge invariance.

 We now coming to the next topic of this talk: hybrids
and their role in hadron dynamics. We start with the hybrid
Hamiltonian and spectrum. This topic in the framework of FCM was
considered in \cite{9,10} The Hamiltonian $H_0$ for hybrid looks
like \cite{1},\cite{9},\cite{10}
\be
H_0^{(hyb)} =\frac{m^2_1}{2\mu_1} + \frac{m^2_2}{2\mu_2} +
\frac{\mu_1+\mu_2+\mu_g}{2} + \frac{\vep^2_\xi+\vep^2_\eta}{2\mu}
+\sigma\sum^2_{i=1} |\ver_g-\ver_i| + H_{str}.\label{22} \ee

Here $\vep_\xi, \vep_\eta$ are Jacobi momenta of the 3-body
system, $H_{self}$ is the same as for meson, while $H_{spin}$ and
$H_{Coul}$ have different structure \cite{10}; $H_{str}$ is the
string term similar to $H_{string}$ in (10).

The main feature of the present approach based on the Backgroung
Perturbation Theory (BPTh), is that valence gluon in the hybrid is
situated at some point on the string connecting quark and
antiquark, and the gluon creates a kink on the string so that two
pieces of the string move independently (however connected at the
point of gluon). This differs strongly from the flux-tube model
where hybrid is associated with the string excitation  as a whole.

 Results for light and heavy exotic
$1^{-+}$ hybrids  also given in \cite{1} and are in agreement with
lattice calculations. Typically an additional gluon in the exotic
$(L=1)$ state "weights" 1.2$\div$1.5 GeV for light to heavy
quarks, while nonexotic gluon $(L=0)$ brings about 1 GeV to the
mass of the total $q\bar q g$ system.

 \section{Hamiltonian and Fock states}

 As was mentioned above the QCD Hamiltonian is introduced in
 correspondence with the chosen hypersurface, which defines internal
 coordinates $\{\xi_k\}$ lying inside the hypersurface,
and the evolution parameter, perpendicular to it. Two extreme
choices are frequently used, 1) the c.m. coordinate system with
the hypersurface $x_4=const.$, which implies that all hadron
constituents have the same (Euclidean) time coordinates
$x_4^{(i)}=const, i=1,...n$, 2) the light-cone coordinate system,
where the role of $x_4$ and $x_4^{(i)}$ is played by the $x_+,
x_+^{(i)}$ components, $x_+=\frac{x_0+x_3}{\sqrt{2}}$.

To describe the structure of the Hamiltonian in general terms
 we first assume that the bound valence states exist for
mesons, glueballs and baryons consisting of minimal number of
constituents. To form the Fock tower of states starting with the
given valence state, one can add gluons and $q\bar q$ pairs
keeping the $J^{PC}$ assignment  intact. At this point we make the
basic simplifying approximation assuming that the number of colors
$N_c$ is tending  to infinity, so that one can do for any physical
quantity an expansion in powers of $1/N_c$. Recent lattice data
confirm a good convergence of this expansion for $N_c=3,4,6$ and
all quantities considered \cite{25} (glueball mass, critical
temperature, topological susceptibility etc.).

Then the construction of the Fock tower is greatly  simplified
since any additional $q\bar q$ pair enters with the coefficient
$1/N_c$ and any additional white (e.g. glueball) component brings
in  the coefficient $1/N_c^2$. In view of this in the leading
order of $1/N_c$ the Fock tower is formed by only creating
additional gluons in the system, i.e. by the hybrid excitation of
the original (valence) system. Thus all Fock tower consists of the
valence component and its hybrid equivalents and each line of this
tower is characterized by the number $n$ of added gluons. Then,
the internal coordinates $\{\xi\}_n$ describe coordinates and
polarizations of $n$ gluons in addition to those of valence
constituents.

We turn now to the Hamiltonian $H$, assuming it to be either the
total QCD Hamiltonian $H_{QCD}$, or the effective Hamiltonian
$H^{(eff)}$, obtained from $H_{QCD}$ by integrating out
short-range degrees of freedom. We shall denote the diagonal
elements of $H$, describing the dynamics of the $n$-th hybrid
excitation of $s$-th valence state ($s=m\{f\bar f\}, gg, 3g,
b\{f_1f_2f_3\})$ for mesons, 2-gluon and 3-gluon glueballs and
baryons respectively with $f_i$ denoting flavour of quarks) as
$H^{(s)}_{nn}$. For nondiagonal elements we need only the lowest
order operators $H^{(s)}_{n,n+1}$ and $H^{(s)}_{n-1, n}$
describing creation or annihilation of one additional gluon, viz.
\be
H_{\bar q qg}=g\int\bar q  (\vex, 0)\hat a (\vex,0 )) q(\vex, 0)
d^3x\label{10a}\ee
\be
H_{g2g} =\frac{g}{2} f^{abc}\int (\partial_\mu
a^a_\nu-\partial_\nu a^a_\mu) a^b_\mu a^c_\nu d^3 x.\label{11a}\ee
As it is clear from (\ref{10a}), (\ref{11a}), the first operator
refers to the gluon creation from the   quark line, while the
second refers to the creation of 2 gluons from the gluon line. In
what follows we shall be mostly interested in the first operator,
which yields dominant contribution at large  energies, and
physically describes addition of   one last cross-piece to the
ladder of gluon exchanges between quark lines, while (\ref{11a})
corresponds in the same ladder to the $\alpha_s$ renormalization
graphs.

The effective Hamiltonian in the one-hadron sector can be written
as follows
\be
\hat H=\hat H^{(0)} +\hat V\label{12a}\ee where $H^{(0)}$ is the
diagonal matrix of operators,
\be
H^{(0)}= \{ H_{00}^{(s)}, H_{11}^{(s)}, H_{22}^{(s)}, ...
\}\label{13a}\ee
 while $\hat V$ is the sum of operators (\ref{10a}) and (\ref{11a}),
 creating and annihilating one gluon. In (\ref{13a}) $H^{(s)}_{nn}$
 is the Hamiltonian operator for what we call the "n-hybrid", i.e.
 a bound state of the system, consisting of $n$ gluons together
 with the particles of  the valence  component. In this way the
 $n$-hybrid for the valence $\rho$-meson is the system consisting
 of $q\bar q$ plus $n$ gluons "sitting" on the string connecting
 $q$ and $\bar q$.

 Before applying the stationary perturbation theory in $\hat V$ to
 the Hamiltonian (\ref{12a}), one should have in mind that there
 are two types of excitations  of the ground state valence Fock
 component: 1) Each of the operators $H_{nn}^{(s)}, n=0,1,...$ has
 infinite amount of excited states, when radial or orbital motion
 of any degree  of freedom is excited, 2) in addition one can add
 a gluon, which means exciting the string and this excitations due
 to the operator $\hat V$ transform the $n-th$ Fock component
 $\psi_n^{(s)}$
 into $\psi^{(s)}_{n+1}$.

 The wave equation for the Fock tower $\Psi_N\{P,\xi\}$ has the
 standard form
 \be
 \hat H\Psi_N=(\hat H^{(0)}+\hat V) \Psi_N=E_N\Psi_N,\label{14a}\ee
where $N$ numerates energy eigenvalues, and $\xi$ is a set of
internal quantum numbers in the $n$-hybrid, or in the integral
form
 \be
 \Psi_N= \Psi_N^{(0)} - G^{(0)} \hat V \Psi_N \label{15a}\ee
 where $G^{(0)}$ is diagonal in Fock components,
 \be
 G^{(0)}(E) =\frac{1}{\hat H^{(0)}-E},~~
 G^{(0)}_{nm}(E) =\delta_{nm} \frac{1}{ H^{(s)}_{nn}
 -E},\label{16a}\ee
 and $\Psi_N^{(0)}$ is the eigenfunction of $\hat H^{(0)}$,
 \be
 \hat H^{(0)} \Psi_N^{(0)}= E^{(0)}_N\Psi_N^{(0)}\label{17a}\ee
 and since $\hat H^{(0)}$ is diagonal, $\Psi^{(0)}_N$ has only one
 Fock component, $\Psi_N^{(0)}=\psi_n(P,\{\xi\}_n),$ $n=0,1,2,...,$
 and the eigenvalues  $E^{(0)}_N$ contain all possible excitation
 energies of the $n$-hybrid, with the number $n$ of gluons in the
 system fixed,
 \be
 E_N^{(0)} = E_n^{(0)} (P)
 =\sqrt{\veP^2+M^2_{n\{k\}}}.\label{18a}\ee
 Here $\{ k\}$ denotes the set of quantum numbers of the excited
 $n$-hybrid.

 From (\ref{15a}) one obtains in the standard way
 corrections to the eigenvalues and eigenfunctions.

 As a first step one should
  specify the  unperturbed functions $\Psi^{(0)}_N$, introducing
  the set of quantum numbers $\{k\}$ defining the excited hybrid
  state for each $n$-hybrid Fock component $\psi_n(P\{\xi\}_n)$; we
  shall denote therefore:
  \be
  \Psi^{(0)}_N=\psi_{n\{k\}}(P,\{\xi\}_n), ~~
  n=0,1,2,...\label{19a}\ee

  The set of functions $\psi_{n{\{k\}}}$ with all possible $n$ and
  $\{k\}$ is a complete set to be used in the  expansion of the
  exact wave-function (Fock tower) $\Psi_N$:
  \be
  \Psi_N=\sum_{m\{k\}}c^N_{m\{k\}}\psi_{m\{k\}}.\label{20a}\ee
  Using the orthonormality  condition
  \be
  \int\psi^+_{m\{k\}}\psi_{n\{p\}} d\Gamma
  =\delta_{mn}\delta_{\{k\}\{p\}}\label{21a}\ee
  where $d\Gamma$ implies integration over all internal
  coordinates and summing over all indices, one obtains from
  (17) an equation for  $c_{m\{k\}}$ and $E_N$,
  \be c^N_{n\{p\}} (E_N-E^{(0)}_{n\{p\}}) = \sum_{m\{k\}} c^N_{m\{
  k\}}V_{n\{p\}, m\{k\}}\label{22a}
  \ee
  where we have defined
  \be
  V_{n\{p\}, m\{k\}}= \int \psi^+_{n\{p\}}\hat V \psi_{m\{k\}}
  d\Gamma.\label{23a}\ee
and $E^{(0)}_{n\{p\}}$ is the eigenvalue of the wave function
component $\psi_{n\{p\}}$.

Consider now the Fock tower built on the valence component
$\psi_{\nu\{\kappa\}}$, where $\nu $ can be any integer. For
$\nu\{\kappa\}= 0\{0\}$ this valence component corresponds to the
unperturbed hadron with minimal number of valence particles. For
higher values of $\nu\{\kappa\}$ the Fock component
$\psi_{\nu\{\kappa\}}$ corresponds to the hybrid  with $\nu$
gluons which after taking into account the interaction is "dressed
up" and acquires all other Fock components, so that the number $N$
in (\ref{20a}) contains the "bare number" $\nu\{\kappa\}$ as its
part $N=\nu\{\kappa\},...$ (at least for small perturbation $\hat
V$).

One can impose on $\Psi_N$ the orthonormality condition
\be
\int\Psi^+_N\Psi_M d\Gamma = \sum_{m\{k\}} c^N_{m\{k\}}
c^M_{m\{k\}} = \delta_{NM}.\label{24} \ee

Expanding  in powers of $\hat V$, one has
\be
c_{m\{k\}}^{N(\nu\{\kappa\})}= \delta_{m\nu}
\delta_{\{k\}\{\kappa\}}+ c^{N(1)}_{m\{k\}}+
c^{N(2)}_{m\{k\}}+...\label{25}\ee
\be
E_{N(\nu\{\kappa\})}=E^{(0)}_{\nu\{\kappa\}}+
 E^{(1)}_N+E^{(2)}_N+...\label{26}\ee

 It is easy to see that $E^{(1)}_N\equiv 0$, while for $c^{(1)}$
 one obtains from (\ref{22a}) the standard expression
 \be
 c^{N(1)}_{n\{p\}}=
 \frac{V_{n\{p\},\nu\{\kappa\}}}{E^{(0)}_{\nu\{\kappa\}}-E^{(0)}_{n\{p\}}}.\label{27}\ee
 In what follows we shall be  interested in  the high Fock
 components, $\nu+l,\{k\}$, obtained by adding $l$ gluons to the
 valence component $\nu\{\kappa\}$. Using (\ref{22a}) and
 (\ref{25}) one obtains
$$
 C^{N(
\nu\{\kappa\})}_{\nu+l,\{k\}}= \sum_{\{k_1\}...\{k_l\}}
\frac{V_{\nu+l\{k\},\nu+l-1\{k_1\}}}{E^{(0)}_{\nu\{\kappa\}} -
E^{(0)}_{\nu+l\{k\}}}\frac{V_{\nu+l-1\{k_1\},\nu+l-2\{k_2\}}}{E^{(0)}_{\nu\{\kappa\}}
- E^{(0)}_{\nu+l-1\{k_1\}}}...$$ \be
\frac
{V_{\nu+1\{k_l\},\nu\{\kappa\}}}{E^{(0)}_{\nu\{\kappa\}} -
E^{(0)}_{\nu+1\{k_l\}}}+O(V^{l+2}).\label{28}\ee

Since $\hat V$ is proportional to $g$, one obtains in (\ref{25})
the perturbation series in powers of $\alpha_s$ for $c^N$ and
hence for $\Psi_N$ (\ref{20a}). One should note that
$\alpha_s(Q^2)$ is the background coupling constant, having the
property of saturation for positive $Q^2$ \cite{8,26} and the
background perturbation series has no Landau ghost pole and is
defined in all Euclidean region of $Q^2$.

The estimate of the mixing between meson and hybrid was done
earlier in the framework of the  potential model for the meson in
\cite{27}. In \cite{21} the mixing between hybrid, meson and
glueball states was calculated in the framework of the present
formalism and we shortly summarize the results. One must estimate
the matrix element (\ref{23a}) between meson and hybrid wave
functions taking the operator $\hat V$ in the form of (\ref{10a}),
where the operator of gluon emission at the point $(\vex, 0)$ can
be approximated as
 $$
 a_\mu(\vex, t)
=\sum_{\vek,\lambda}\frac{1}{\sqrt{2\mu(\vek)V}}\times $$
\be
\times [\exp({i\vek\cdot \vex-i\mu t)}e^{(\lambda)}_\mu c_\lambda
(\vek)+e^{(\lambda)}_\mu c^+_\lambda(\vek) \exp({-i\vek\cdot
\vex+i\mu t})]
 \label{42}
 \ee

 Omitting for simplicity all polarization vectors and
 spin-coupling coefficients which are  of the order of unity, one
 has the matrix element
 \be
 V_{Mh} = \frac{g}{\sqrt{2\mu_g}} \int \varphi_M (\ver
 )^\mu\psi_h^+ (0, \ver) d^3r \label{43}\ee
 where $\varphi_M(\ver), ~~^\mu\psi_h(\ver_1,\ver_2)$ are meson
 and hybrid wave functions respectively, and in (\ref{43}) it is
 taken into account that the gluon is emitted (absorbed) from the
 quark position.

 Using realistic Gaussian approximation for the wave functions in
 (\ref{43}) one obtains the estimate \cite{21}
 \be
 V_{Mh}\approx g \cdot 0.08~{\rm GeV}.\label{44}\ee
 A similar estimate is obtained in \cite{21} for the
 hybrid-glueball mixing matrix element, while the meson-glueball
 mixing is second-order in (\ref{44}).

 Hence the  fist-order hybrid admixture coefficient (\ref{27}) for the meson
 is
 \be
 C_{Mh} =\frac{V_{Mh}}{E^{(0)}_M-E^{(0)}_h} =\frac{V_{Mh}}{\Delta
 M_{Mh}}\label{45}\ee
 and for the ground state low-lying mesons when $\Delta M_{Mh
 }\sim 1$ GeV it is small, $C_{Mh}\sim 0.1-0.15$, yielding a 1-2\%
 probability. However for higher states in the region $M_M\ga 1.5$
 GeV, the mass difference $\Delta M_{Mh}$ of mesons and hybrids
 with the same quantum numbers can be around 200 MeV,  and the
 mixing becomes extremely important, also for meson-glueball
 mixing, which can be written as
 \be
 C_{MG} =\sum_h \frac{V_{Mh} V_{hG}}{\Delta M_{Mh} \Delta
 M_{hG}}\label{46}\ee
 and $V_{Mh} \sim V_{hG}$. A good example is given by the three
 $f_0$ mesons: $f_0(1390), f_0(1500)$ and  $f_0(1710)$ studied on
 the lattice in \cite{28}.

  The authors \cite{28} arrive at the following result
 of careful lattice studies:
 $$|f_0(1710)>=0.859|g>+0.302|s\bar s>+0.413|n\bar n>, $$
 $$|f_0(1500)>=-0.128|g>+0.908|s\bar s>-0.399|n\bar n>,
 $$
\be
 |f_0(1390)>=-0.495|g>+0.290|s\bar s>+0.819|n\bar n>.
\label{100} \ee

From (\ref{100}) it is clear that a strong mixing occurs between
states in the region 1.4--1.7 GeV, however the dominant valence
component in all three cases is clearly visible: it is $(n\bar n)$
for  $f_0(1390)$, $(s\bar s)$ for $f_0(1500)$ and the two-gluon
glueball ($g$) for $f_0(1710)$.

\section{Sea quarks in the  spectrum and strong decays}

In all discussion above the dynamics in the $q\bar q$ system was
described by the Wilson  loop $W_\sigma(C)$ in (\ref{1}). However
the formalism in (\ref{1})-(\ref{3}) is correct only in the large
$N_c$ limit, which  holds for the lower part of the spectrum with
accuracy of the order of 10\%. For higher excited states with the
radius exceeding 1.3-1.4 fm one should take into account the
admixture of quark pairs. This can be accomplished formally
replacing $W_\sigma(C)$ in (\ref{1}) by the product $W_\sigma (C)
\prod_f \det (m_f+\hat D)$. Using the heat kernel
(Fock-Feynman-Schwinger) representation for the determinant one
has  $$ Re~\ln \det (m+\hat D) =\frac12\ln\det (m^2+\hat
D^2)=$$\be =\int d^4 x\int^\infty_0\frac{ds}{s}
 \exp (-m^2 s) (Dz)_{xx} \exp  (-K) W(C_x).\label{38b}\ee
where $W(C)$ is the Wilson loop without spin factors, present in
(2), and  $C_x$ is the closed contour beginning and ending at the
point $x$.
 In this way the determinant can be expanded as a series
in powers
 of number  of sea-quark loops and the averaging  yields
\be
 \lan W_\sigma (C) \det (m+\hat D)\ran =
 = \lan W_\sigma(C)\ran
\lan \det (m+\hat D)\ran +\frac{a_1}{N_c} W_1(C,
C_x)+...\label{39b}\ee
 where
 \be
 W_1(C,C_x)\equiv \lan W_\sigma(C)W(C_x)\ran - \lan W_\sigma
 (C)\ran \lan W(C_x)\ran. \label{40}\ee
It was shown in \cite{29} that the interaction of Wilson
 loops given by (\ref{40}) effectively produces holes in the
 original world sheet of the string in $W_\sigma(C)$. It was argued
 in \cite{30}, that for unstable states with the life-time $T\sim
 1$fm and average radius larger than 1.4fm the holes due to the
 $q\bar q$ pairs can be in   metastable equilibrium which can be
 called "the predecay state". In this state the linear potential
 between quarks starts to  saturate, and the resulting meson
 masses calculated in \cite{30} are strongly decreased by 200-500
 MeV, which is in good  agreement with experimental data. The
 resulting radial Regge trajectories are of surprisingly good
 linear form, and are given in \cite{31}.

 Another phenomenon, where sea quarks play the crucial role is the
 strong decay process. As it was shown in \cite{32} one can
 distinguish three possible mechanisms of strong decays, where
 nonperturbative QCD is involved: 1) chiral mechanism, which
 appears after bosonization \cite{33}, \cite{34} 2) string
 breaking mechanism and 3) pair creation via the intermediate
 hybrid or glueball formation.

Taking into account valence (perturbative) gluon field $a_\mu(x)$,
one has after averaging over background gluon field and
bosonization the following effective quark-meson Lagrangian
\cite{32}-\cite{34} $$ L_{QML} =\int d^4 x\int d^4y
\{~^f\psi^+_{a\alpha}(x)\left [[ i(\hat
\partial - ig\hat a) +im_f]_{\alpha\beta}\delta^{(4)}(x-y)
\delta_{fg}+ \right .$$
\be
\left .+ iM_S\hat U^{(fg)}_{\alpha\beta} (x,y) \right
]^g\psi_{a\beta} (y)- 2n_f [J(x,y)]^{-1} M^2_S(x,y)\}\label{41}\ee
where $\hat U=\exp (i\hat \phi(x,y) \gamma_5)$, and $\hat
\phi=t^a\phi_a/f_\pi$ is the pionic field, $n_f$ is number of
flavours, and $J(x,y)$ is the kernel proportional to the integral
of field correlator, $f,g$ -- flavour indices, while $M_S(x,y)$ is
the effective quark mass operator containing the string connecting
quark to the closest antiquark position, for details and notations
see \cite{32}-\cite{34}.

All decay mechanisms listed above are contained in the Lagrangian
(\ref{41}), e.g. the chiral mechanism obtains by expanding $\hat
U(\phi)$ in powers of $\phi_a$,  as for the details and comparison
with experiment, see \cite{33} and the literature cited in there.

The nonperturbative string breaking mechanism is described in
(\ref{41}) by the term $\psi^+ M_S\psi$, where it is essential
that the mass operator  $M_S(x,y)=M_S^{(br)}$ enters at the vertex
of the quark-antiquark formation (rather than the string mass
operator in the quark  propagation). The explicit computation of
$M_S^{(br)}$ in the string breaking term was done in \cite{32},
and yields
\be
\Delta L^{(br)} =\frac{2T_g\sigma}{\sqrt{\pi}}\int d^4 x \bar \psi
(x) \psi (x).\label{42b}\ee

As it is clear from (\ref{42b}) one obtains the $^3P_0$ mechanism
with the predicted coefficient, which agrees with the generic
phenomenologically fitted value.

Finally, the mechanism 3) is given by the combination of the
$q\bar q$ generation by the perturbative gluon, due to the term
$\psi^+\hat  a\psi$, with the subsequent string world sheet
enveloping  the gluon and quark trajectories. In this way one
obtains  the Hybrid- Mediated Decay (HMD) and -- for  the OZI
violating processes -- the Glueball-Mediated Decay (GMD),
suggested in \cite{32}. The relative role and theoretical
estimates of all three mechanisms was not elaborated in \cite{32}.

\section{Conclusions}

It was shown above that the Vacuum Correlator Method ({VCM)  is a
powerful tool for the investigation of all effects in the QCD
spectrum. In particular the low-lying part of the spectrum, where
the string breaking is not essential, is well described by the
 Hamiltonian (\ref{15}) containing the minimal number of
input parameters: current quark masses, string tension and
$\alpha_s$. Here effects of mixing and decay are inessential (less
than 10\%) and the leading large $N_c$ approximation is valid.
However for higher masses (and for high-energy processes) one
should take into account the   Fock-tower  structure of hadrons,
where in particular hybrids contribute significantly.
Superficially these Fock towers are similar to the light-cone Fock
states considered in \cite{35}, however the (essential) difference
is that in our case all Fock components are bound states -- mesons
or hybrids in the leading large $N_c$ limit, while in \cite{35}
only perturbative dynamics is present.

It was demonstrated above that also the $1/N_c$ effects, namely
glueball admixture in mesons and decay amplitudes, are easily
computed in the framework of the VCM, and moreover no new
parameters are introduced. However only first step is done in
\cite{32} for the construction of decay and production amplitudes.

The author is grateful to K.G.Boreskov, A.B.Kaidalov and
O.V.Kancheli for useful discussions.

The partial support of the INTAS grants 00-110 and 00-366 is
gratefully acknowledged.

This work was supported by the Federal Program of the Russian
Ministry of Industry, science and Technology No 40.052.1.1.1112.

\end{document}